\documentclass[aps,prl,preprint,superscriptaddress,showpacs,floatfix]{revtex4}
\usepackage{graphicx}

\begin{document}

%\preprint{}

\title{Optical Anisotropy of Electronic Excitations in Elliptical Quantum Dots}

\author{Achintya Singha}
\email{ a.singha@sns.it} \affiliation{NEST INFM-CNR and Scuola
Normale Superiore, Pisa 56126, Italy}
\author{Vittorio Pellegrini}
\affiliation{NEST INFM-CNR and Scuola Normale Superiore, Pisa 56126, Italy}
\author{Sokratis Kalliakos}
\affiliation{Department of Materials Science and Technology,
University of Crete, P.O. Box 2208, 71003 Heraklion Crete, Greece}
\author{Biswajit Karmakar}
\affiliation{NEST INFM-CNR and Scuola Normale Superiore, Pisa 56126, Italy}
\author{Aron Pinczuk}
\affiliation{Depts of Appl. Phys \& Appl. Math. and of Physics, Columbia University, New York 10027, USA}
\affiliation{Bell Laboratories, Alcatel-Lucent, Murray Hill, New Jersey 07974, USA}
\author{Loren N. Pfeiffer}
\affiliation{Bell Laboratories, Alcatel-Lucent, Murray Hill, New Jersey 07974, USA}
\author{Ken W. West}
\affiliation{Bell Laboratories, Alcatel-Lucent, Murray Hill, New Jersey 07974, USA}


%\author{}
%\thanks{}
%\affiliation{}


\date{\today}

\begin{abstract}
The authors report that anisotropic confining potentials in laterally-coupled semiconductor quantum dots (QDs) have large impacts in optical transitions and energies of inter-shell collective electronic excitations. The observed anisotropies are revealed by inelastic light scattering as a function of the in-plane direction of light polarization and can be finely controlled by modifying the geometrical shape of the QDs. These experiments show that the tuning of the QD confinement potential offers a powerful method to manipulate electronic states and far-infrared inter-shell optical transitions in quantum dots.
\end{abstract}


\maketitle

Electronic states in circular quantum dots (QDs) are characterized by rotational symmetry that gives rise to an atomic-like shell structure in which states with opposite angular momentum are degenerate \cite{reim}.  A convenient description of single-particle QD levels is provided by the Fock-Darwin (FD) orbitals with energies given by $E_{nm} = \hbar \omega _o (2n+m+1)$ = $\hbar \omega _o (N+1)$, where n=0,1.., m=0,1,... are the radial and azimuthal quantum numbers, respectively, and $\omega _o$ is the in-plane confinement energy. The studies of the spin and charge configurations in QDs by transport and optical techniques have been a major research theme in modern condensed matter with several implications in the field of quantum computation \cite{kouw1,hanson,optics}. QDs in the low-density limit are also fascinating nanostructures. In this limit, the breakdown of the FD scheme leads to correlated configurations that reveal the impact of fundamental electron interaction \cite{kall}.
\par
The shell structure of circular QDs manifests in the spectra of neutral collective modes probed by inelastic light scattering \cite{shuller}. The light scattering experiments are able to probe both spin and charge inter-shell monopole excitations with even change of $\Delta N$ in the FD shell and without change in angular momentum, as required by light scattering selection rules. In case of charge inter-shell excitations, the selection rules dictates that the polarizations of the incoming and scattered photons have to be parallel. In circular QDs, these electronic charge excitations are isotropic in the plane i.e. their energy does not depend on the direction of the polarizations of the incoming and scattered photons, which reflects the rotational symmetry of the nanostructure.
\par
In this letter we demonstrate that asymmetric QDs with few electrons present a peculiar optical anisotropy that manifests in their inter-shell excitation spectra. By nanofabrication we designed different AlGaAs/GaAs quantum dot structures composed by two closely-spaced QDs having the overall lateral shape like the one shown in Fig.1. The two main parameters characterizing our nanostructures are the diameter $2R$ of the two QDs and the inter-dot distance $D$. The lateral anisotropy is gradually modified by increasing the value of $D$.
\par
We measured the low lying collective inter-shell charge
excitations with polarizations along and perpendicular to the
inter-dot axis by resonant inelastic light scattering. The spectra
reveals a large energy-splitting between the excitations in the
two configurations. The evolution of energy-splitting in the two
polarization configurations in samples with different $D$ values
represent a direct proof of the breakdown of rotation invariance
and consequent lifting of the degeneracies of the single-particle
levels present in a circularly-shaped dot. In addition, the
observed anisotropy in the scattered light implies the breakdown
of the light scattering selection rules that could be induced by
heavy-light hole mixing due to the asymmetric confinement. We
remark that the impact of ellipsoidal deformation on the Coulomb
blockade transport properties has been studied as a function of a
magnetic field \cite{tarucha} but, to the best of our knowledge,
no direct experiments probing the optical anisotropy of electronic
excitations has been reported so far. This is particularly
striking since ground states, selection rules and matrix elements
of both dipole and  multipole transitions of elliptically-shaped
QDs have been extensively calculated \cite{marlo,cantele,
serra,madhav,sun,xu,li} as well as valence band mixing effects and
interband optical properties \cite{cortez}. In the single electron
picture, the conduction-band energy level structure of the
elliptical QD is given by $E_{n_{X}n_{Y}} = \hbar \omega _X
(n_{X}+1/2)+ \hbar \omega _Y (n_{Y}+1/2)$ \cite{madhav} where
$\omega_X$ and $\omega_Y$ represents the confinement energies in
the $X$ and $Y$ directions respectively, as shown in Fig.1
(right-upper panel). Studies of photoluminescence polarization
anisotropies have been also reported particularly in nanorod
crystals \cite{alivisatos}.
\par
Samples were fabricated from a 25 nm wide, one-side
modulation-doped Al$_{0.1}$Ga$_{0.9}$As/GaAs quantum well with
measured low-temperature electron density $n_e$ = 1.1$
\times$10$^{11}$ cm$^{-2}$ and a mobility of 2.7$\times$10$^6$
cm$^{2}$/Vs. QDs were produced by inductive coupled plasma
reactive ion etching. QD arrays (with sizes 100 $\times$ 100 $\mu
$m containing $10^4$ single identical QD replica) were defined by
electron beam lithography. Deep etching (below the doping layer)
was then achieved. The lateral asymmetry is introduced by
fabrication of two identical quantum dots, with identical radii
$R$, close to each other having centre to centre distance $D$.
When the distance  $D$ is less than $2R$ the confinement resembles
that of a single quantum dot with an elliptical-like potential
(see Fig.1). As D increases towards values $D>2R$, the two quantum
dots tend to decouple into two isolated circular QDs (see Fig.3).
Here we focus on QDs with $R$ = 90 nm and values of $D$ in the
range between 170 nm and 260 nm. Thanks to the presence of a
depletion layer, the effective confinement (green shaded region in
Fig.1) is less than the geometrical values. In the uncoupled
regime at large $D$ the estimated population of each QDs is 3-5
electrons and $\omega _Y \approx $ 4 meV \cite{garcia}.
\par
The experiments were performed in a backscattering configuration
at temperatures T =1.9 K. A tuneable ring-etalon Ti:Sapphire laser
was focused on a 100 $\mu$m-diameter area for excitation of
the quantum dots array and the scattered light from the quantum
dots array having the same polarization direction of the incoming
light was collected through a series of optics, dispersed by a
triple-grating spectrometer and detected by a CCD multi-channel
detector. A polarization rotator between the laser and the sample
and a $\lambda/2$ plate in front of the spectrometer were used to
define the two polarization configurations: XX-both incident
and scattered light are parallel to the inter dot axis and YY-the
same for perpendicular direction.
\par
The right-bottom panel in Fig.1 shows representative inelastic
light scattering spectra of inter-shell charge excitations in the
two polarization configurations. The spectra reveal two
distinct excitations (A and B) that are selectively observed in
the two polarization configurations YY and XX, respectively. The
thick blue and red lines are Gaussian fits to the experimental
data. The observed energy difference between the peak A and B is
the signature of the large optical anisotropy due to the asymmetry
of the in-plane confinement potential.
\par
The manifestation of the optical anisotropy in light scattering is intriguing. In a cylindrical QD the electronic states can be labelled by their angular momentum quantum numbers. Therefore interband transitions involving both light-hole and heavy-hole states should be isotropic in the X-Y plane i.e. do not depend on the polarization direction in X-Y plane. The deviation from this behavior in our QDs could be explained by invoking mixing between heavy- and light-hole states induced by the external asymmetric potential. While a quantitative calculation of this effect is beyond the scope of this paper, we note that the origin of optical anisotropies in asymmetric nanostructures and the role of valence-band mixing have been theoretically addressed in several works \cite{besombes,krebs,cortez,sheng}.
\begin{figure}
\begin{center}
\includegraphics*[width=8cm]{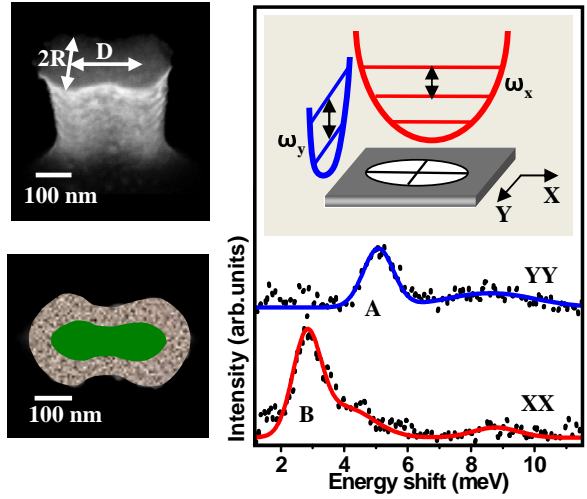}
\end{center}
\caption{SEM images of coupled quantum dot. The semi-elliptical green shaded area represents the effective confinement region of the electrons. This coupled system clearly shows the anisotropy of the confinement potential. XX represents the configuration of charge excitation when both incident and scattered light are parallel to the inter dot axis. YY represents the same when they are perpendicular to the inter dot axis. The energy parabola along the major and minor axes are included and the respective excitations A and B, are marked. Inelastic light scattering charge excitation spectra for the light polarization configuration XX and YY. The spectra are shifted vertically for clarity. }
\end{figure}
\par
The energy splitting of the modes A and B that is around 2 meV, can be related to $\omega _Y-\omega _X $ although their absolute energies are expected to deviate from $\omega _Y$ and $\omega _X$ due to the impact of electron correlations that we expect significant in these regimes of low electron occupation \cite{garcia}. The splitting between the inelastic light scattering resonance enhancement peaks in the XX and YY configurations shown in Fig.2 is also close to 2 meV. This fact suggests small splittings of valence band states of the X and Y inter-band transitions involved in the light scattering process. If we consider $\omega _Y-\omega _X \approx 2$ meV, and $\omega _Y \approx 4 meV$ \cite{garcia} we deduce $\omega _X\approx 2$ meV and therefore an ellipticity factor $\delta = \omega _X/\omega _Y = 0.5$ for the QD shown in the inset to Fig.1.
\par
\begin{figure}
\begin{center}
\includegraphics*[width=6.5cm]{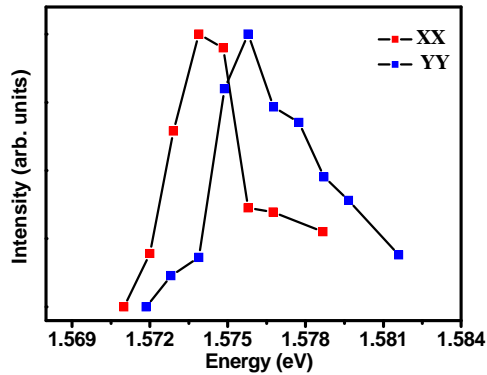}
\end{center}
\caption{Inelastic light scattering resonant enhancement profiles of the inter-shell charge collective modes in the two polarization configurations.}
\end{figure}
\par
To demonstrate tuning of the optical anisotropy we report in Fig.3 the evolution of the charge excitation spectra in both XX and YY configurations as a function of the inter-dot distance $D$.
\begin{figure}
\begin{center}
\includegraphics[width=6.5cm]{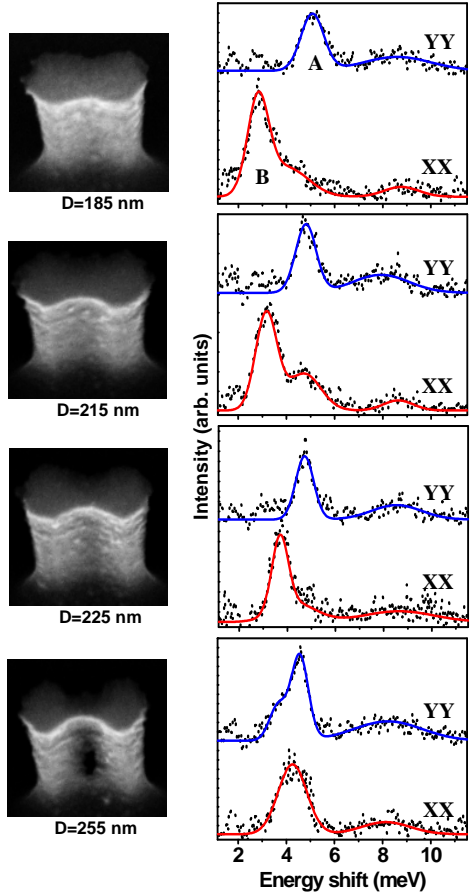}
\end{center}
\caption{Left vertical panels: SEM images of the coupled QDs with different inter-dot distance $D$. Right vertical panels: charge excitation spectra for XX and YY polarization configuration.}
\end{figure}
The results displayed in Fig.3 indicate that the energy difference
between peaks A and B decreases with the increase of D and, as
expected, it vanishes in the case of nearly decoupled dots when $D
> 2R$. In this case the re-established circular symmetry leads to
similar charge spectra in the two polarizations. Finally Fig.4
reports the A-B energy splitting values as a function of inter-dot
distance. The red line is an exponential fit to the data with the
function $A(e^{-D}-e^{-D_o})$ where $D_o \approx 270 $ nm.
\begin{figure}
\begin{center}
\includegraphics[width=6.5cm]{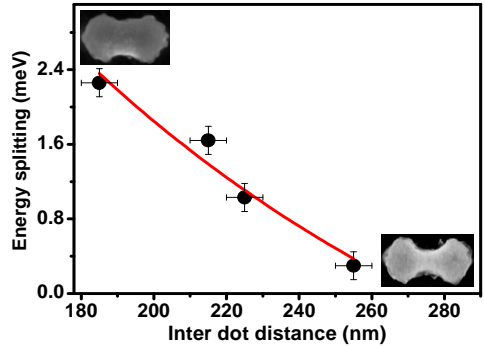}
\end{center}
\caption{Evolution of the energy splitting between the two
chargemodes (A-B) in YY and XX polarization configurations as a
function of the inter-dot distance. The red line is an exponential
fit to the data. The insets show SEM images of the QDs
corresponding to the two extreme points.}
\end{figure}
\par
In conclusion, we have shown that the tuning of the QD confinement potential offer a method to manipulate electronic states with great precision. The breakdown of rotational invariance here reported might offer new venues for the exploration of electron correlation effects in these nanostructures. In addition, the anisotropic QDs here studied might be usable for polarization-dependent detection of far-infrared radiation.
\begin{acknowledgments}
This work was supported by the projects MIUR-FIRB No. RBIN04EY74,
A. P.  is supported by the National Science Foundation under
Grants No. DMR-0352738, DMR-0803691 and CHE-0641523. By the
Department of Energy under Grant No. DE-AIO2-04ER46133, and by a
research grant from the W. M. Keck Foundation. V.P acknowledge
partial support of the Italian Academy and NSEC at Columbia
University. We acknowledge useful discussion with Massimo Rontani,
Guidi Goldoni and Elisa Molinari.
\end{acknowledgments}


\end{document}